\documentclass[a4paper,fleqn,usenatbib]{mnras}

\usepackage{mathptmx}
\usepackage[T1]{fontenc}
\usepackage{ae,aecompl}

\usepackage{graphicx}	% Including figure files
\usepackage{amsmath}	% Advanced maths commands
\usepackage{amssymb}
\usepackage{subfigure}% Extra maths symbols

\def\nii{[N\,{\sc ii}]}

\def\kms{km\,s$^{-1}$}

\title[Outflows from the nucleus of IRASF23199+0123]{Gemini IFU, VLA and HST observations of the OH Megamaser galaxy IRASF23199+0123: the hidden monster and its outflow}

\author[C. Hekatelyne et al.]{C. Hekatelyne,$^{1}$\thanks{E-mail: hekatelyne.carpes@gmail.com}
Rogemar A. Riffel,$^{1}$ Dinalva Sales,$^2$  Andrew Robinson,$^3$  
\newauthor Jack Gallimore,$^4$ Thaisa Storchi-Bergmann,$^5$ Preeti Kharb,$^6$ Christopher O'Dea,$^{7,8}$ 
\newauthor  Stefi Baum,$^{7,9}$
\\
$^{1}$ Departamento de F\'isica, CCNE, Universidade Federal de Santa Maria, 97105-900, Santa Maria, RS, Brazil\\
$^{2}$ Instituto de Matem\'atica, Estat\'istica e F\'isica, Universidade Federal do Rio Grande, Rio Grande 96203-900, Brazil \\
$^{3}$ School of Physics and Astronomy, Rochester Institute of Technology, 84 Lomb Memorial Drive, Rochester, NY 14623, USA \\
$^4$ Department of Physics, Bucknell University, Lewisburg, PA 17837, USA \\
$^5$ Departamento de Astronomia, Universidade Federal do Rio Grande do Sul. 9500 Bento Gon\c{c}alves, Porto Alegre, 91501-970, Brazil \\
$^6$ National Centre for Radio Astrophysics, Tata Institute of Fundamental Research, S. P. Pune University Campus, Post Bag 3,\\ Ganeshkhind, Pune 411 007, India \\
$^7$ Department of Physics and Astronomy, University of Manitoba, Winnipeg, MB, R3T 2N2, Canada\\
$^8$ School of Physics \& Astronomy, Rochester Institute of Technology, 84 Lomb Memorial Dr., Rochester, NY 14623, USA.\\
$^9$ Center for Imaging Science, Rochester Institute of Technology, 84 Lomb Memorial Dr., Rochester, NY 14623, USA.\\}

\date{Accepted XXX. Received YYY; in original form ZZZ}

\pubyear{2017}

\begin{document}
\label{firstpage}
\pagerange{\pageref{firstpage}--\pageref{lastpage}}
\maketitle

\begin{abstract}
We present Gemini Multi-Object Spectrograph (GMOS) Integral field Unit (IFU), Very Large Array (VLA) and Hubble Space Telescope (HST) observations of the OH Megamaser (OHM) galaxy IRASF23199+0123.
Our observations show that this system is an interacting pair, with two OHM sources associated to the eastern (IRAS\,23199E) member.
 The two members of the pair present somewhat extended radio emission at 3 and 20~cm, with flux peaks at each nucleus. The GMOS-IFU observations cover the inner $\sim$6\,kpc of IRAS23199E at a spatial resolution of 2.3~kpc. The GMOS-IFU  flux distributions in H$\alpha$ and [N\,{\sc ii}]$\lambda$6583 are similar to that of an HST [N\,{\sc ii}]+H$\alpha$ narrow-band image, being more extended along the northeast-southwest direction, as also observed in  the continuum HST F814W image. The GMOS-IFU H$\alpha$ flux map of IRAS\,23199E shows three extranuclear knots attributed to star-forming complexes. We have discovered a Seyfert\,1 nucleus in this galaxy, as  
 its nuclear spectrum shows an unresolved broad (FWHM$\approx$2\,170\,km\,s$^{-1}$) double-peaked H$\alpha$ component,  from which we derive a black hole mass of M$_{BH}$= 3.8$^{+0.3}_{-0.2}\times 10^{6}$\,M$_{\odot}$. The gas kinematics shows low velocity dispersions ($\sigma$) and low [N\,{\sc ii}]/H$\alpha$ ratios for the star-forming complexes and higher $\sigma$ and [N\,{\sc ii}]/H$\alpha$ surrounding the radio emission region, supporting interaction between the radio-plasma and ambient gas. The two OH masers detected in IRASF23199E are observed in the vicinity of these enhanced $\sigma$ regions, supporting their association with the active nucleus and its interaction with the surrounding gas. The gas velocity field can be partially reproduced by rotation in a disk, with residuals along the north-south direction being tentatively attributed to emission from the front walls of a bipolar outflow.
\end{abstract}

%galaxies: ISM; galaxies:ULIRGs; galaxies: dynamics; galaxies: individual (IRAS23199+...)
%%%%%%%%%%%%%%%%%%%%%%%%%%%%%%%%%%%%%%%%%%%%%%%%%%

%%%%%%%%%%%%%%%%% BODY OF PAPER %%%%%%%%%%%%%%%%%%

%\section{Rogemar's comments}

%\begin{itemize}
 %   \item Colocar a Figura 1, junto com a imagem do HST e campo do GMOS. Arrumar escala de fluxo para unidades fisicas (e nao arbitrarias) e colocar a unidade do eixo x
  %  \item Mapas de fluxo podem/devem ser em escala logaritmica. Serah que nao conseguimos melhores ajustes para o [N\,{\sc ii}]+Ha? Coloca o escript de medidas.pro, profit6g, profit3g e cubo no dropbox. Quando eu tiver um tempo, eu brinco um pouco.
   % \item colocar pontos do WHAN para as duas componentes separadas
%    \item estudar como podemos levar em conta a contribuição em absorção do Halpha e Hbeta.
%\end{itemize}

\section{Introduction}

(Ultra)luminous infrared galaxies ([U]LIRGs) are among the most luminous objects in the universe showing infrared (IR) luminosities of L$_{IR}> 10^{12}$ L$_{\odot}$. These objects are believed to represent a key stage in the evolution process of galaxies in which tidal torques associated with mergers drive gas into the galaxy core, leading to the feeding/triggering of nuclear starbursts or the fuelling of embedded active galactic nuclei \citep[AGN - e.g.][]{Dinalva2015}.

These merging systems provide a conducive environment for OH maser emission and approximately 20\% of [U]LIRGs contain extremely luminous OH masers, emitting primarily in the 1667 and 1665 MHz lines with luminosities 10$^{2-4}$ L$_\odot$ \citep{Lo2005,Darling2002}.
%OH megamasers are luminous masers and tracers of star formation and merging galaxies (and possibly massive black holes) that do not suffer from dust or atmospheric extinction and provides accurate redshifts (Briggs 1998; Townsend et al. 2001; Darling \& Giovanelli 2002c; Darling \& Giovanelli 2006).
The OH megamasers (OHMs) are commonly associated to merging systems, but the environment that produces this phenomenon is still not completely understood. Many OHM hosts present a composite spectrum, showing both AGN and starburst features. A possible explanation for these features is that they originate in a central AGN, contaminated by emission of circum-nuclear star-forming regions, as the sampling of the observations usually corresponds to more than one kpc at the galaxies. Alternatively, the OHM galaxies could represent a transition stage between a starburst and the eruption of an AGN, as suggested by \citet{Darling2006}.

%The origin of the intense infrared emission is still not well established. The two phenomena have been  pointed as the main mechanisms of production of the IR emission from the nucleus of these galaxies: the presence of an AGN or intense star formation. {\bf *** outra ref aqui -- tentar discutir sobre como investigar estes dois fenomenos ***}

%OH maser emission is commonly observed associated to [U]LIRGs and some of these merging systems are responsible to produce strong maser emission, called OH megamasers (OHMs). {\bf *** a definição de megamaser abaixo estah confusa - tentar reescrever - sugiro usar os comandos citep e citet para as referencias.***} 

Considering the above scenario, it becomes relevant to investigate the nature of the gas ionization source of OH megamaser galaxies. In this paper, we present Gemini Multi-Object Spectrograph (GMOS) Integral Field Unit (IFU) observations, VLA continuum data and Hubble Space Telescope (HST) narrow and broad band images of the galaxy IRASF23199+0123, which is an interacting pair of ULIRGs that presents OH megamaser emission.
 This galaxy is part of a sample of 15 OH Megamaser galaxies, for which we have already HST images from a project that has the overall goal  of relating the merger state and OH maser properties to AGN and Starburst nuclear activity.  We have selected targets for IFU observations from the 15 galaxies observed with HST, on the basis of the morphology revealed by the images.  The present paper is a pilot study based on multiwavelength observations, aimed to study the gas kinematics and excitation of one OH Megamaser galaxy and that we hope to extend to the whole sample. 

IRASF23199+0123 has a redshift $z=0.13569$ \citep{Darling2006}, corresponding to a distance of 558~Mpc for which 1$^{\prime\prime}$ corresponds to 2.7 kpc at the galaxy, assuming a Hubble constant of $H_0=73$ \kms\,Mpc$^{-1}$.
Its OH maser emission was first detected in the Arecibo survey, that observed 52 objects with $0.1<z < 0.26$ \citep{Darling2001}. \citet{Darling2006} used spectroscopic data obtained with the Palomar 5 m telescope with the Double Spectrograph in order to perform an optical spectroscopic study of the properties of the sample of the Arecibo survey and identified the nuclear emission of IRASF23199+0123 as being due to a Seyfert 2 nucleus, based on emission-line ratios.

Our GMOS-IFU data comprise observations of the central region of the eastern galaxy of the  IRASF23199+0123 pair and our aim is to map the distribution and kinematics of the optical line emitting gas and investigate the excitation mechanism of the nuclear emission. This is the first time that OH megamaser galaxies have been observed with an Integral Field Spectrograph,  allowing a two-dimensional look at the gas excitation and kinematics in detail. This paper is organized as follows. Section 2 describes the observations and data reduction procedure and section 3 explains the emission-line fitting process and present maps for the emission-line flux distributions and kinematics, as well as the HST and VLA radio continuum images, while in section 4 the results are discussed. Finally, the conclusions of this work are presented in section 5.
%This is a simple template for authors to write new MNRAS papers.
%See \texttt{mnras\_sample.tex} for a more complex example, and \texttt{mnras\_guide.tex}
%for a full user guide.

%All papers should start with an Introduction section, which sets the work
%in context, cites relevant earlier studies in the field by \citet{Others2013},
%and describes the problem the authors aim to solve \citep[e.g.][]{Author2012}.

\section{Observations and data reduction}

\subsection{VLA Radio Continuum data}

We observed IRASF23199+0123 with the Karl G. Jansky Very Large Array (VLA) on Apr 20, 2014. The observations included X-band (8--10~GHz) continuum, L-band (1--2~GHz) continuum, and L-band spectral line observations of the redshifted OH (1665/1667~MHz) maser lines. 
The L-band and X-band observations comprised respectively three and one 10-minutes scans, alternating with 3-minute scans of the phase calibrator, J2320+0513. We observed the source 3C48 in both X-band and L-band for flux and bandpass calibration.

The VLA~pipeline in CASA \citep{McMullin2007}, was used for data reduction. This includes initial data flagging and phase, flux and bandpass calibrations. The continuum images were generated using multi-frequency synthesis \citep[e.g.][]{Conway90,Rau2011} with natural weighting and deconvolved using the Cotton-Schwab variant of the CLEAN algorithm \citep{Schwab84}. Imaging included simultaneous deconvolution of neighbouring radio sources within the primary beam. We applied three rounds of phase-only self-calibration based on CLEAN models for the radio continuum (self-calibration is reviewed by \citet{Pearson84}). For the L-band continuum image, the restoring beam is 1\farcs69 $\times$ 1\farcs28, PA~14$^\circ$, and the background {\it rms}  is 0.024~mJy~beam$^{-1}$. The restoring beam of the X-band continuum image is 0\farcs339$\times$0\farcs260, PA~29$^\circ$, and the background rms is 0.0093~mJy~beam$^{-1}$. The radio continuum images are presented in Figure~\ref{fig:radiocontours}.

The spectral line visibilities were continuum subtracted in two steps. First, we produced CLEAN continuum models based on line-free channels, and the CLEAN models were subtracted from the observed visibilities. Second, we removed any residual continuum using the CASA task {\it uvcontsub}; the continuum was determined by averaging visibility spectra over line-free channels. The continuum-subtracted, spectral line data cube was produced using standard Fourier inversion and CLEAN
deconvolution. The expected line width of the 1667~MHz feature is 0.68~MHz \citep{Darling2001}; therefore, to improve the signal-to-noise, we binned the spectral line data to 0.23~MHz channels (roughly $1/3$ line width). The restoring beam of the OH spectral line cube is 1\farcs77 $\times$ 1\farcs36, PA~14$^\circ$, and the typical background {\it rms}  for a single 0.23~MHz channel is 0.44~mJy~beam$^{-1}$.

\subsection{Hubble Space Telescope data}

HST images of IRASF23199+0123 were acquired using the Advanced Camera for Surveys (ACS) with the broad-band  filter F814W, the narrow-band filter FR656N and  medium-band FR914M filter as part of a snapshot survey program to observe a sample of OHMGs (Program id 11604; PI: D.J. Axon). The total integration time was 600 s in the broad band (I) F814W filter, 200 s in the medium-band filter and 600 s in the narrow band H$\alpha$ FR656N filter. The bandpass of the FR656N narrow band filter includes both H$\alpha$ and [N{\sc ii}]$\lambda 6548, 83$ emission lines.
We have processed the final images in order to remove cosmic rays using IRAF task $lacos_{im}$ \citep{Vandokkum2001}. 
The continuum free H$\alpha+$[N{\sc ii}] image of IRASF23199+0123 was constructed according to the following the steps:
(i) the count rates of a few stars were obtained in both the medium (FR914M) and narrow band (FR656N) ramp filter images; (ii) from the count rate ratios the mean scaling factor was computed and applied to the medium band FR914M image; (iii) the scaled FR914M image was then subtracted from the narrow-band FR656N image. We next visually inspected our continuum subtracted H$\alpha+$[N{\sc ii}] image to assure that the residual fluxes of foreground stars were negligible within the uncertainties.
This procedure results in typical flux uncertainties of 5-10\%  \citep[see][]{Hoopes1999,Rossa2000,Rossa2003}. 

The HST images (see Fig.~\ref{fig:hst-images}) show for the first time that IRASF23199+0123 is indeed an interacting pair and we have obtained IFS of the eastern member of the par (hereafter IRAS23199E).

\subsection{GMOS IFU data}

IRAS23199E was observed using the GMOS \citep{hook04} IFU \citep{allington-smith02} as part of the program GS-2013B-Q-86 (PI: D. Sales). Only the eastern nucleus was observed, as it presents a steep continuum flux distribution and with bright guide stars available in the GMOS patrol field. The  major axis of the IFU was oriented along position angle $PA=215^\circ$, approximately along the major axis of the galaxy. The total on source exposure time was 4800 s\ divided into 4 individual exposures of 1200 s each. The observations were performed on August 29, 2013  using the B600 grating with the IFU in the one slit mode, in combination with the GG455 filter. This setup resulted in an angular coverage of 5\farcs0$\times$3\farcs5, covering the spectral region from 450\,nm to 750\,nm at a spectral resolution of 1.7 \AA, as obtained from the measurement of the full width at half maximum (FWHM) of typical emission lines of the Ar lamp spectrum used for the wavelength calibration. 

The data reduction process was performed using routines of the GEMINI package in the Image Reduction and Analysis Facility \citep[IRAF,][]{tody86,tody93} software and followed the standard procedure of spectroscopic data reduction \citep{lena2014}. First, we subtracted the bias level from each image, performed flat-fielding  and trimming. Then, we applied wavelength calibration to the data using the spectra of arc lamps as references and subtracted the sky emission. Finally, we performed flux calibration using a sensitivity function obtained from the spectrum of the H600 photometric standard star observed during the same night of the object observations. 

Finally, datacubes for each individual exposures were created at an angular sampling of 0\farcs1$\times$0\farcs1, which were then median combined using the IRAF {\it gemcombine} task to obtain the final datacube for the object. The peak of the continuum emission was used as a reference during the mosaicking of the individual datacubes and we used the {\it avsigclip} algorithm for bad pixel/cosmic ray removal. We estimated the angular resolution as 0\farcs85 from the measurement of the FWHM of the flux distribution of field stars present in the acquisition image of the galaxy. This angular resolution corresponds to $\sim$2.3 kpc at the galaxy for its adopted distance ($d=558~$Mpc).

In order to remove noise from the final datacube, we performed a spatial filtering using a Butterworth bandpass filter \citep{gonzalez02,menezes14,menezes15} via the IDL routine $bandpass_-filter.pro$\footnote{The routine is available at $https://www.harrisgeospatial.com/docs/bandpass_-filter.html$}, which allows the choice of the cut-off frequency ($\nu$) and the order of the filter $n$. A low value of $n$ (e.g. 1) is close to a Gaussian filter, while a high value (e.g. 10) corresponds to an ideal filter. We used $n=5$ and $\nu=0.15$~Ny, chosen by comparing the filtered cube with the original one. For lower values of $\nu$, besides the removal of spatial noise, the filter excludes also emission from the galaxy, while for larger values of $\nu$ the filtering procedure is not efficient. The filtering process does not change the angular resolution of the data and all measurements presented in the forthcoming sections were done using the filtered cube.

\section{Results}

\subsection{H$\alpha$+[N\,{\sc ii}] GMOS-IFU spectra}
 
In order to map line fluxes, line-of-sight velocity ($V_{LOS}$) and velocity dispersion ($\sigma$) of the emitting gas, we fitted the emission-line profiles of H$\alpha$ and \nii$\lambda\lambda$6548,6583 with Gaussian curves. The fitting was performed using modified versions of the line-PROfile FITting ({\sc profit}) routine \citep{profit}, which provides as outputs the emission-line flux, the centroid velocity, the velocity dispersion and their corresponding uncertainties. Besides the [N\,{\sc ii}]+H$\alpha$ emission lines, the nuclear spectrum includes also the [O\,{\sc i}]$\lambda$6300.3 emission line, which shows a similar profile to that of the [N\,{\sc ii}] and H$\alpha$ narrow components, being well reproduced by a single Gaussian curve.  
%Although the spectral range covers the blue part of the spectra, the signal-to-noise ratio at the blue is too small to allow the detecton of other emission lines.

% and \sii$\lambda\lambda$6716,6730

The fitting process for the H$\alpha$ and [N\,{\sc ii}]$\lambda\lambda$6548,6583 emission lines was performed simultaneously. 
We noticed that in the inner 0\farcs8 radius (corresponding to about two resolution elements), the fitting of a single Gaussian component for each emission-line does not reproduce the observed profile, while at larger distances from the nucleus a single component provides a good fit. Thus, at locations farther then 0\farcs8 from the nucleus (defined as the location of the peak of the continuum emission) we fitted the H$\alpha$+\nii\ complex using one Gaussian per line, adopting the following constraints: (i) we kept tied the kinematics of the [N\,{\sc ii}] lines, such that the two lines have the same velocity and velocity dispersion, and (ii) fixed the [N\,{\sc ii}]$\lambda6583$/[N\,{\sc ii}]$\lambda6548$ intensity ratio to its theoretical value (3). The underlying continuum was fitted by a linear function. 

Within 0\farcs8 from the nucleus, we tested two possibilities for the fit to the H$\alpha$+\nii\ complex. In the first, we fitted the line profiles using four Gaussian curves, in order to include a broad component to represent H$\alpha$. The resulting fit does not reproduce the observed profiles adequately.
%The second attempt was done using two Gaussian curves for each emission line (total of six components), which also provided a good fit for the profiles. Each line was fitted by a narrow central component and a blueshifted broad component. 
In the second fit, the \nii\ lines were fitted by a single Gaussian component, while the H$\alpha$ profile was fitted by the same narrow component plus two broad components. This procedure resulted in a better fit to the profiles. As the presence of two broad components is restricted to the nucleus and the corresponding emission is not resolved, we propose that they actually represent a single double-peaked line originating in the Broad Line Region (BLR). Such double-peaked components are not uncommon in AGN \citep[e.g.][]{sb17}. 

%The first one is the use of two gaussians for each emission line of the complex [N\,{\sc ii}]+H$\alpha$. In this scenario, a blueshifted broad component (FWHM 700 kms$^{-1}$) is presented in all lines, which could be due to gas in outflows from the nucleus of the galaxy. In the second scenario, the [N\,{\sc ii}] lines are fitted by a single gaussian curve, while the H$\alpha$ profile is fitted by one narrow component plus two broad components. In this case, the origin of the broad H$\alpha$ components would be due to a double peak emission of the Broad Line Region.
%Considering that the H$\alpha$+\nii\ complex can be fitted by both six or five components, resulting in a good representation of the observed profiles within the inner 0\farcs8, we have chosen the model with five components because the flux distribution in the broad component is not resolved, as pointed out above, supporting an origin in the BLR. 

\begin{figure}
	\includegraphics[width=\columnwidth]{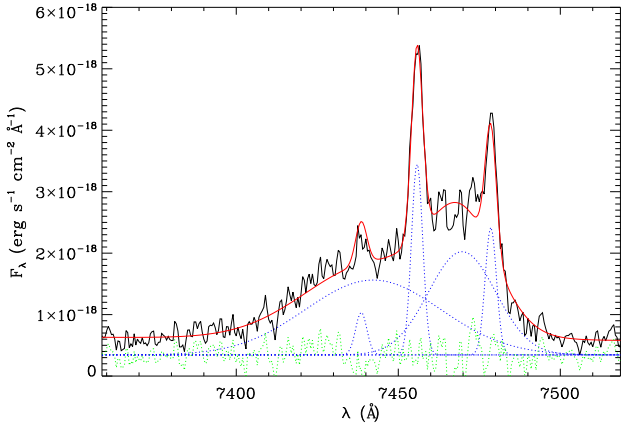}
    \caption{Fit of the nuclear spectrum (within 0\farcs8 from the nucleus)  of IRAS23199E  comprising the H$\alpha$+\nii\ complex}. The observed profile is shown in black, while the blue dotted lines represent the broad and narrow components. The red line is the result of the fit and the green dotted line shows the residual of the fit plus an arbitrary constant.
    \label{fig:ajuste5g}
\end{figure}

In Figure~\ref{fig:ajuste5g} we show the resulting fit of the nuclear spectrum, where the observed profiles are shown in black, the best fit model in red and the individual components as dotted blue lines. As the emission of the BLR is not resolved, the width and central wavelength of each broad component as well as their relative fluxes were kept fixed for all spaxels, while their amplitudes were allowed to vary to enable for smearing by the seeing. 
The values of the centroid velocities and velocity dispersions were obtained by fitting an integrated spectrum of the inner 0\farcs8.  The centroid velocities relative to the systemic velocity are $-$416 \kms and 668 \kms  for the blueshifted and redshifted components, respectively.  The corresponding velocity dispersions are $\sigma=955$~\kms\ for the blue component and $\sigma=475$~\kms\ for the red component. In the hypothesis described above that these two components represent a single double-peaked line, this line profile has a full-width at half maximum of 2\,170 km\,s$^{-1}$. 
The systemic velocity adopted in this paper is $v_{s}$=37\,947$\pm$2 km s$^{-1}$ (corrected for the heliocentric rest frame), as derived by the modelling of the gas velocity field as discussed in Sec.~\ref{gas_kin}. 

%We choose the five gaussians fitting possibility to model the emitter gas and one can see an example of fitting in figure \ref{fig:ajuste5g}. It occurs because the data are not resolved, presenting spatial resolution of 0.88. As the Broad Line Region is not resolved, the width and central wavelenght of each broad component were kept fixed and the amplitude of each Gaussian was allowed to fluctuate.

%Moreover the $\chi^{2}$ values for the six gaussians fit and the five gaussians fit are almost the same. So, we decided to minimize the free parameters. As the Broad Line Region is not resolved, the width and central wavelenght of each broad component were kept fixed and the amplitude of each Gaussian was allowed to fluctuate.

\subsection{OH maser spectra}

\begin{figure}
\includegraphics[width=1.05\columnwidth]{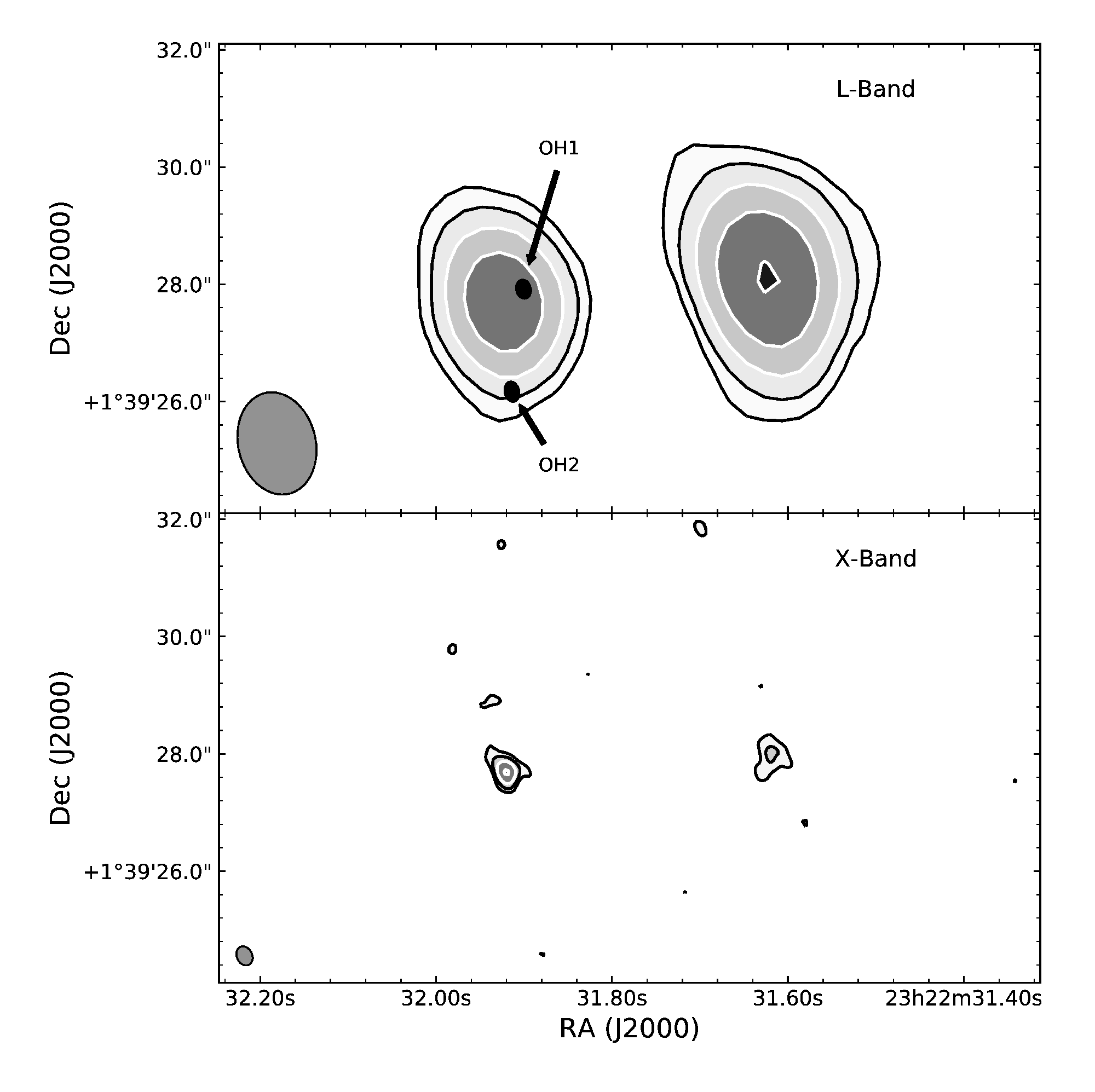}
\caption{Top panel: VLA L-Band (1.6~GHz) continuum image of IRASF23199+0123, shown as filled contours. The contours are (black) 0.071 ($3\sigma$), 0.15, (white) 0.32, 0.69, \& 1.5~mJy beam$^{-1}$. The OH1 and OH2 labels identify the locations where the OH maser sources were detected (see Fig.~\ref{fig-ohspectra}). Bottom panel: VLA X-Band (8~GHz) continuum image of IRASF23199+0123, shown as filled contours. The contours are (black) 0.0278 ($3\sigma$), 0.0647, (white) 0.150, \& 0.349~mJy~beam$^{-1}$.}
\label{fig:radiocontours}
\end{figure}

\begin{figure}
\includegraphics[width=\columnwidth]{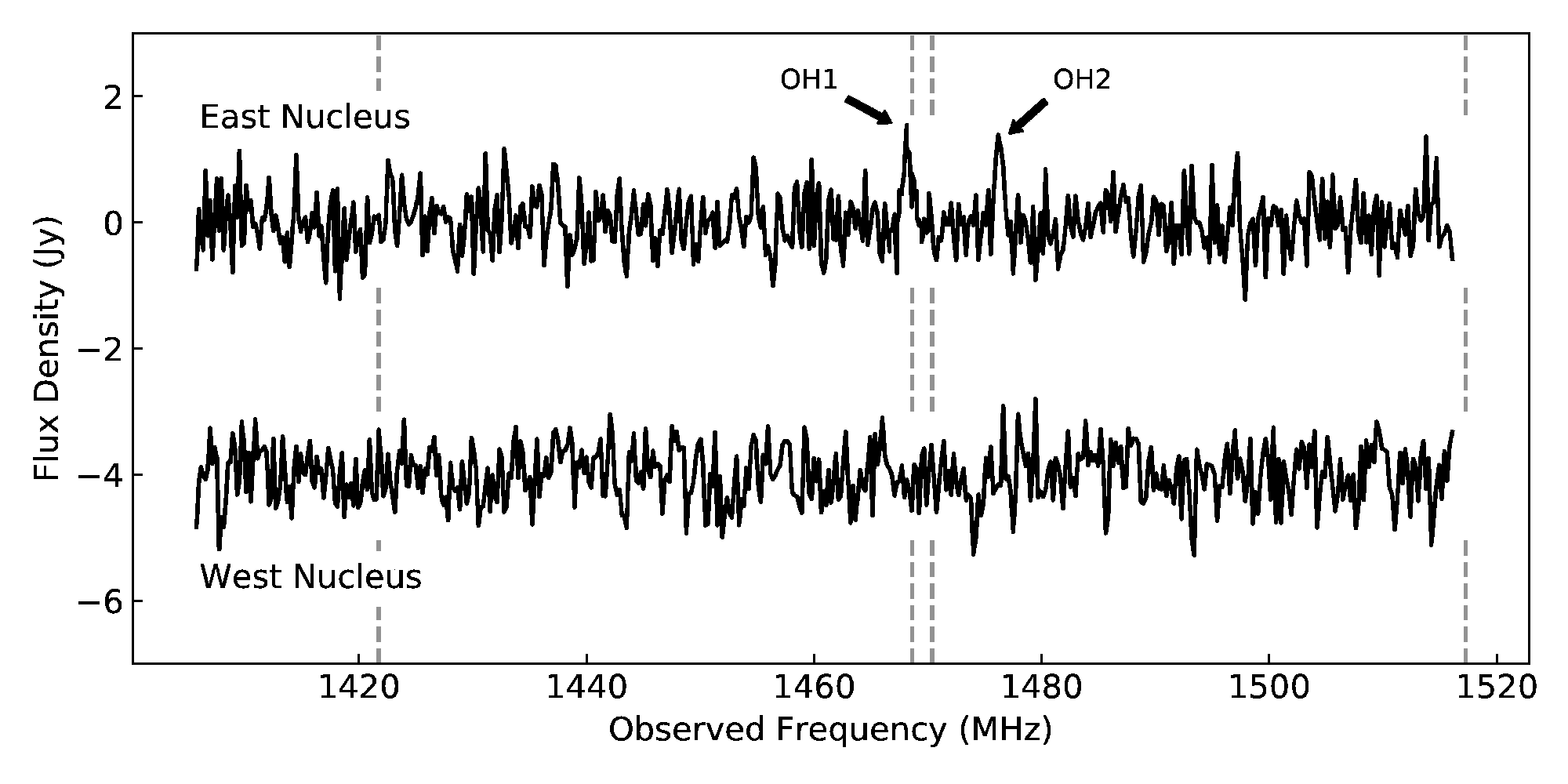}
\caption{VLA OH maser spectra of the eastern (top) and western (bottom, offset by -4~mJy) nuclei. The vertical, dashed grey lines mark the expected, redshifted frequencies for the 1612~MHz, 1665~MHz, 1667~MHz, and 1712~MHz maser features. Two spectral features are detected at the position of the eastern nucleus, marked OH1 and OH2. OH1 was originally detected by \citet{Darling2001}; OH2 is a new detection.}
\label{fig-ohspectra}
\end{figure}

Figure~\ref{fig-ohspectra} shows the VLA OH spectra extracted at the positions of the eastern (IRAS23199E)  and western nuclei. OH masers are detected only towards the eastern nucleus. Two line features appear in the spectrum, which we label `OH1', detected at $5.4\sigma$ significance, and `OH2', a $4.1\sigma$ detection. To within the measurement uncertainties, OH1 matches the central frequency and peak flux density of the maser feature detected by \citet{Darling2001}. To our knowledge, OH2 has never been detected before; OH2 falls outside the bandpass of the \citet{Darling2001} Arecibo observations. The position of OH1 and OH2 defections are identified in the top panel of Fig.~\ref{fig:radiocontours}.

Fig.~\ref{fig-ohspectra} shows that OH1 appears to identify with the 1665~MHz line at the redshift of the eastern nucleus.  Assuming OH1 is indeed the 1665~MHz line, its centroid velocity would be $v =+85\pm28$~\kms\ relative to systemic in the rest frame of the eastern nucleus. OH2 is however significantly offset from the expected heliocentric frequencies of possible emission lines. It may identify either with 1665~MHz line at  $v = -1557\pm22$~\kms\ or the 1667~MHz line at $v = -436\pm29$~\kms\ relative to systemic.

%%%\begin{figure*}
%%%\includegraphics[width=1.7\columnwidth]{lband-contours-markers.ps}
%%%\caption{VLA L-Band (1.6~GHz) continuum image of IRASF23199+0123, shown as filled contours. The contours are (black) 0.071 ($3\sigma$), 0.15, (white) 0.32, 0.69, \& 1.5~mJy beam$^{-1}$. The OH1 and OH1 labels identify the locations where the OH maser emission (see Fig.~\ref{fig-ohspectra}) were detected.}
%%%\label{fig-lbandcontours-markers}
%%%\end{figure*}

%%%\begin{figure*}
%%%\includegraphics[width=1.7\columnwidth]{xband-contours.ps}
%%%\caption{VLA X-Band (8~GHz) continuum image of IRASF23199+0123, shown as filled contours. The contours are (black) 0.0278, 0.0647, (white) 0.150, \& 0.349~mJy~beam$^{-1}$.  }
%%%  \label{fig-xbandcontours}
%%%\end{figure*}

%%\begin{figure*}
%%	\includegraphics[width=2\columnwidth]{radio.ps}
%%    \caption{VLA radio continuum images of IRASF23199+0123 at 20 cm (left panel) and  3 cm (right panel).}
%%    \label{fig:radio}
%%\end{figure*}

%%Figure~\ref{fig:radio} shows the radio continuum images of IRASF23199+0123 obtained with the VLA at 20~cm (left panel) and 3 cm (right panel). Both nuclei are detected at both frequencies. At 20~cm the eastern nucleus present extended emission to up to 1$\farcs$0 from the nucleus,  while the western nucleus shows extended emission to up to 1\farcs5 from it. Both nuclei show the most extended emission along the north-south direction. 

%%The western nucleus presents only a faint emission at 3~cm, being more elongated along the north-south direction. Strong and compact emission is observed for the eastern nucleus, with extended emission being observed to up to 0\farcs8 to the north-northeast, and to up to 0\farcs5 to the south and west-southeast. Additional faint emission is observed at 1\farcs5 north-northeast of the nucleus. 

%%\textcolor{blue}{Rogemar: I think that one of the radiastronomers can give a better description of the radio images.}

\subsection{HST images}\label{hst-ima}
\begin{figure*}
	\includegraphics[width=1.9\columnwidth]{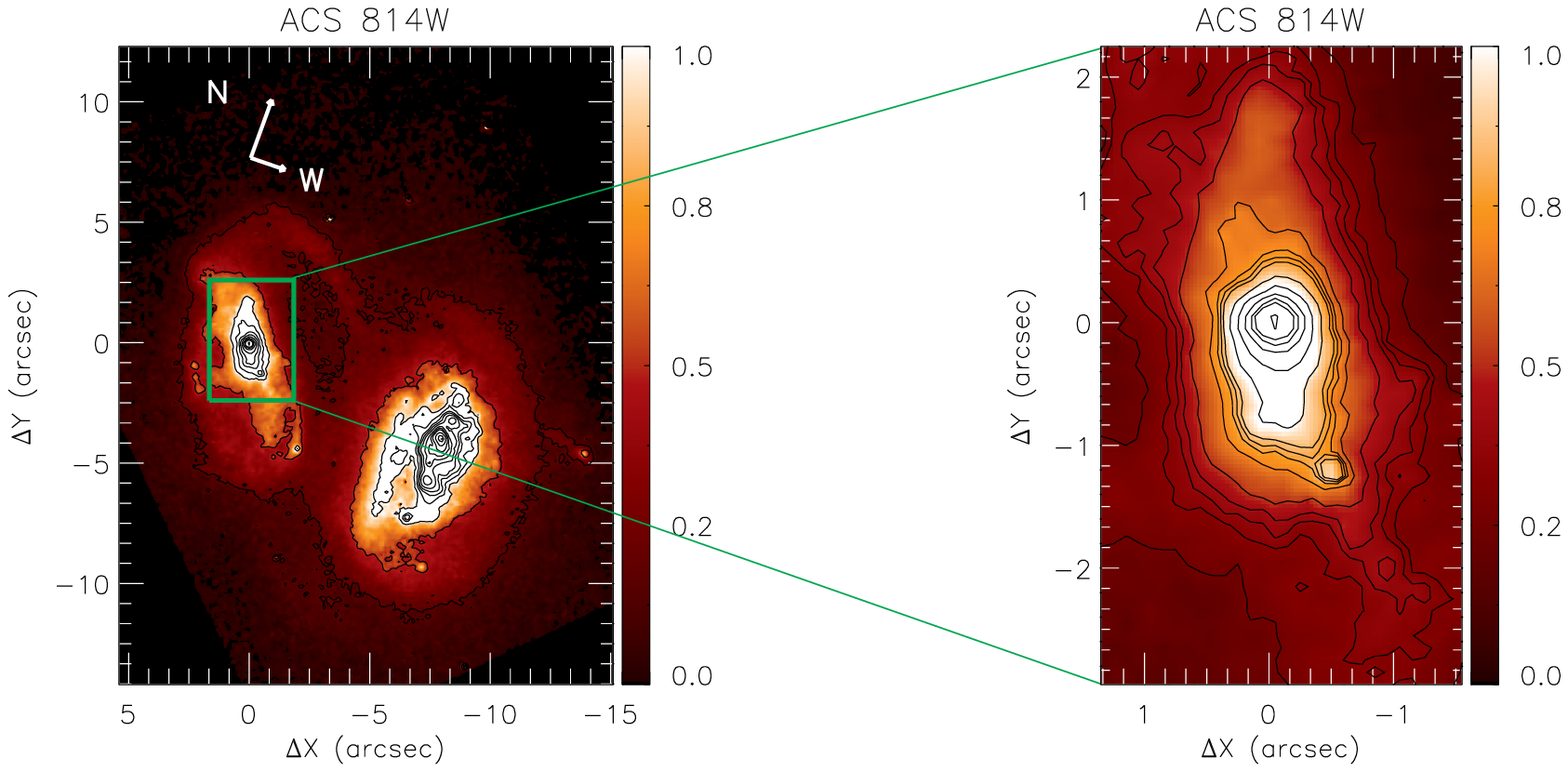}\\
	
	\vspace{0cm}
	
	\includegraphics[width=1.9\columnwidth]{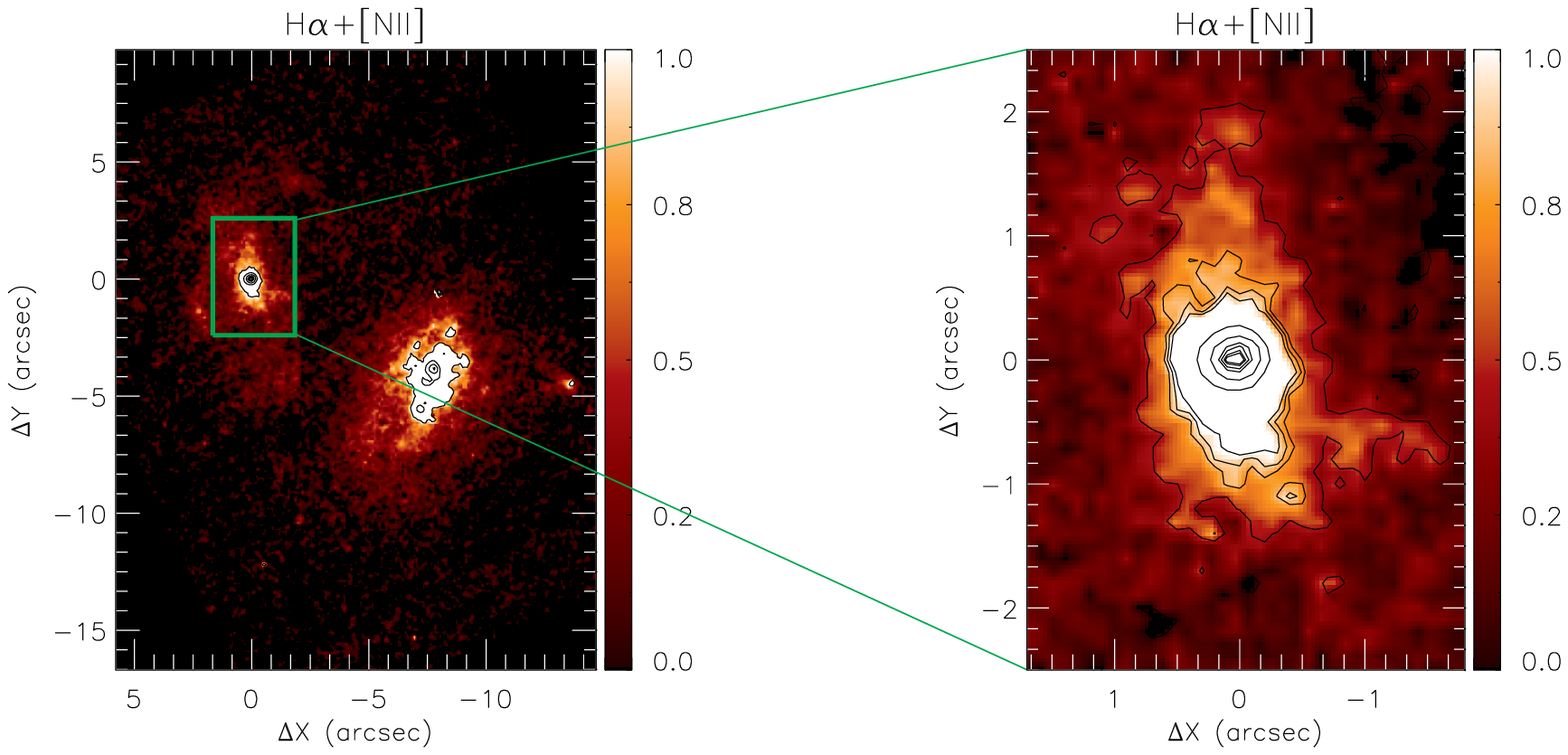}
    \caption{Top panels: Left -- Large-scale image (ACS/HST F814W - i band).  Right -- Zoom of the region observed with GMOS. Bottom panels: Left -- HST large-scale continuum free H$\alpha$+[N\,{\sc ii}] image. Right -- Zoom at the region observed with GMOS IFU. The green boxes show the GMOS IFU field of view (3\farcs5$\times$5\farcs5) and the color bars show the fluxes in arbitrary units.
}
%{\bf Bottom panels: Left -- Structure map. The green box represents the IFU field. Right -- Expanded image of the IFU field.}}
    \label{fig:hst-images}
\end{figure*}

Figure \ref{fig:hst-images} presents the HST broad-band continuum F814W image (top panel) and the narrow-band H$\alpha$+[NII] image (bottom panel) of the inner 20$\times$20 arcsec$^2$ of IRASF23199+0123, which reveal that this system is an interacting pair.  The images in the left panels show both galaxies of the pair, while the right panels show a zoomed in view of IRAS23188E within the same field of view of the GMOS IFU observations. The green box in the left panels corresponds to the field of view of our GMOS data. The HST images were rotated to the same orientation of the GMOS data.

 The F814W continuum image of IRAS23199E presents the highest intensity levels in an elongated structure at $PA\approx45^\circ$ suggesting that the galaxy is highly inclined. 
The western galaxy of the pair shows a less elongated flux distribution, suggesting a more face-on orientation, although the flux distribution towards the center is not uniform, but presents a complex structure. The F814W image  also reveals a structure that resembles a spiral arm to the north that seems to connect the two galaxies, that may be a tidal tail connecting the two galaxies. The linear distance between the galaxy nuclei, projected in the plane of the sky is 24~kpc, assuming a distance to the galaxy of 558~Mpc. This is a lower limit as we do not know the orientation of the plane containing both galaxies or their nuclei relative to the plane of the sky.

%The green box represent the IFU field covered by the observations (3.5'' $\times$ 5'') and on the right, the image shows the expanded IFU field in details. It shows extended emission up to 3 arcsec from the observed nucleus of IRASF23199+0123. 

The bottom left panel of Fig.\,\ref{fig:hst-images} shows the continuum-free  HST H$\alpha$+[N\,{\sc ii}] narrow-band image of the two galaxies. The bottom right panel shows a zoom of the region observed with GMOS-IFU, covering the central part of IRAS23199E. The H$\alpha$+[N\,{\sc ii}] flux distribution is similar to that of the continuum, and suggests the presence of two tidal tails, one to the northeast and another to the southwest of the nucleus.

\subsection{Emission-line flux distributions}

\begin{figure}
\begin{flushleft}
	\includegraphics[width=1.0\columnwidth]{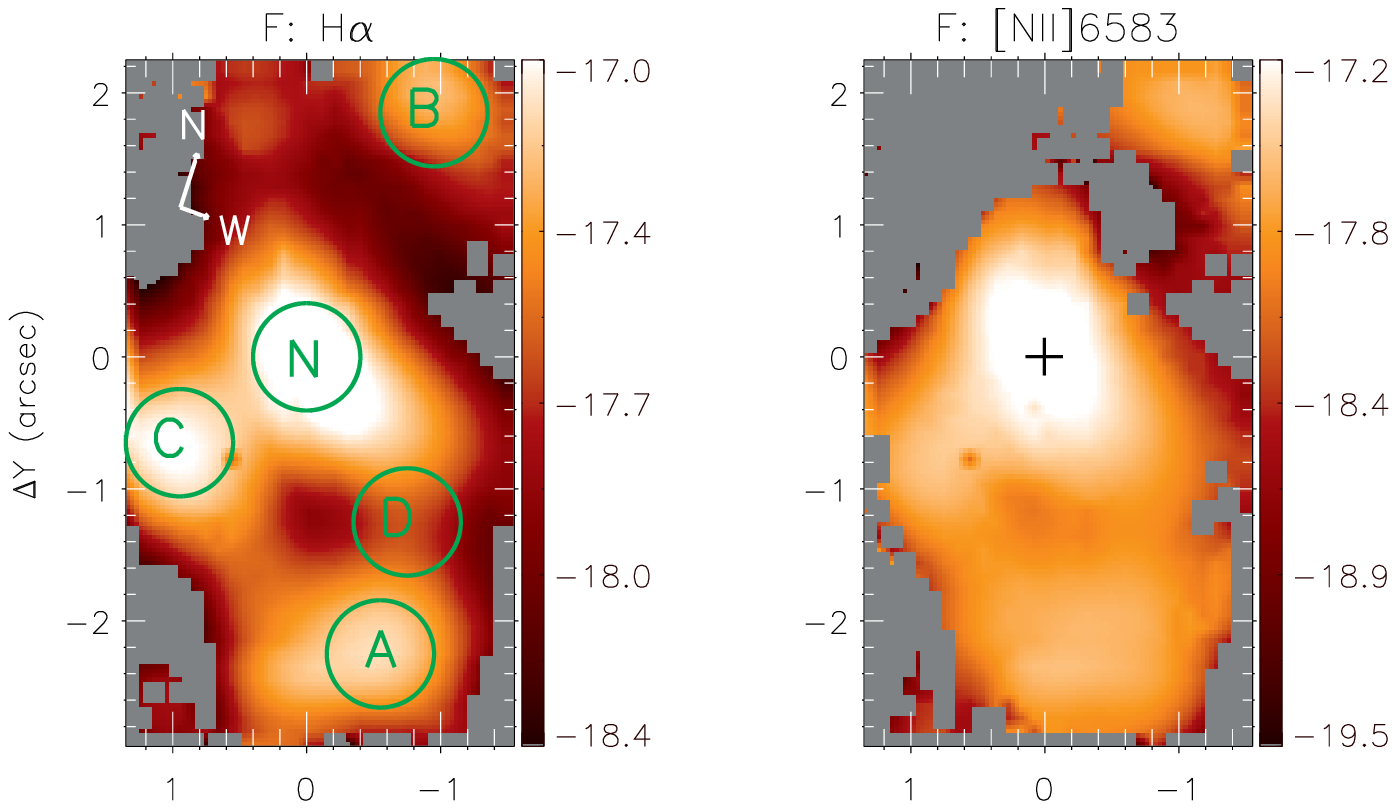}
	
	\vspace{-0.5cm}
	
	\includegraphics[width=1.0\columnwidth]{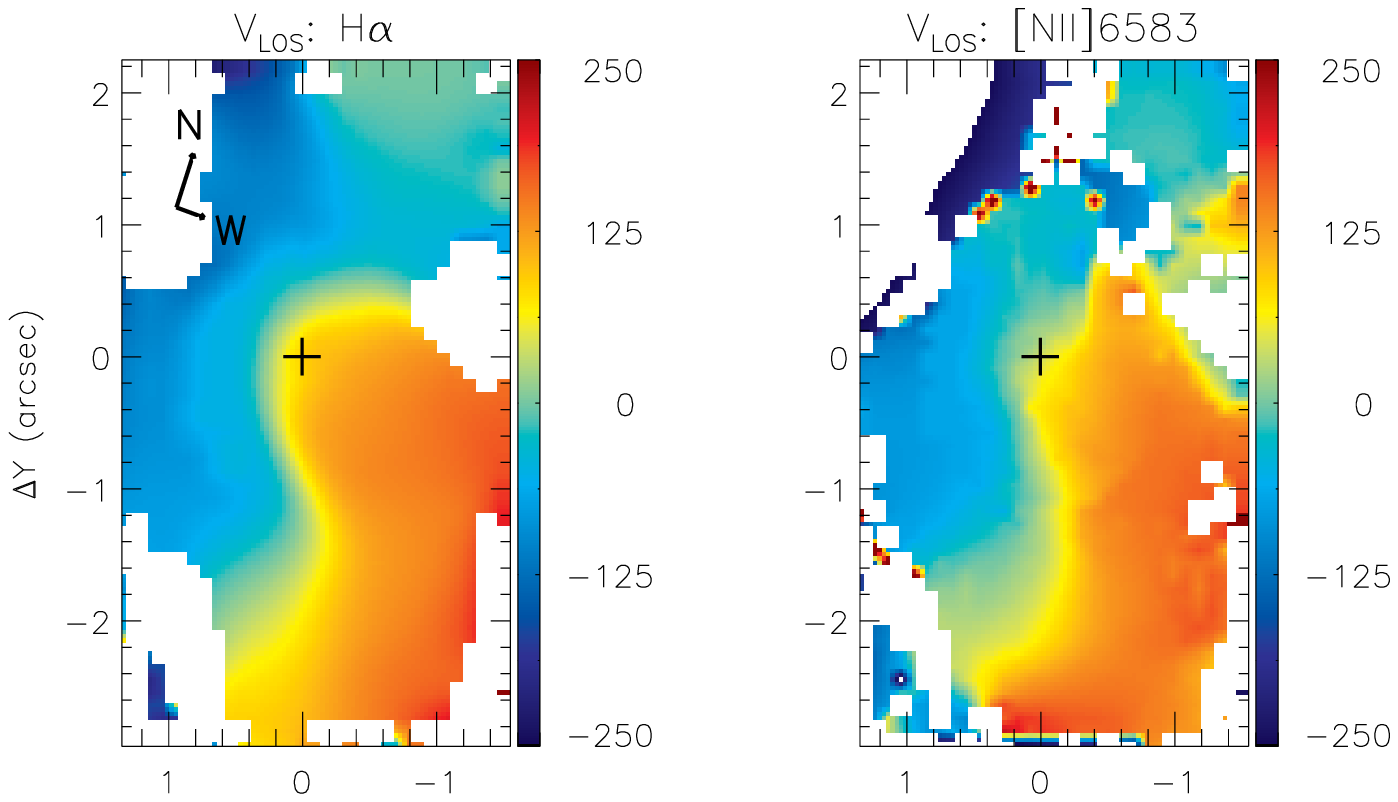}
	
	\vspace{-0.5cm}
	
	\includegraphics[width=1.0\columnwidth]{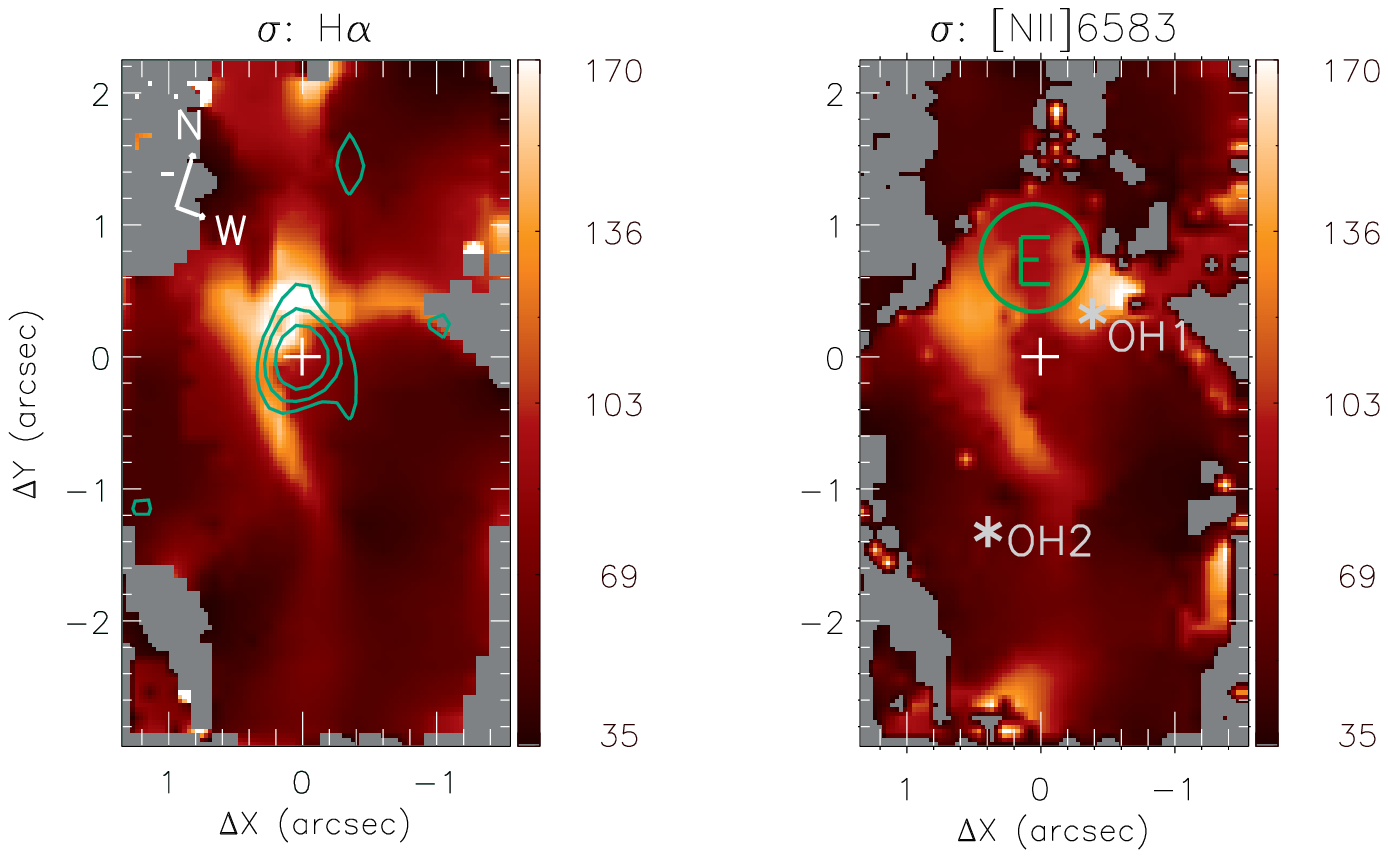}

    \caption{Top panels: flux maps in the H$\alpha$ (left) and [N\,{\sc ii}]$\lambda$6583 (right) emission lines. The color bars show the fluxes in logarithmic units of erg\,s$^{-1}$\,cm$^{-2}$. Central panels: line-of-sight velocity fields for the  H$\alpha$ (left) and [N\,{\sc ii}] (right) emitting gas. The color bars show the velocities in units of km\,s$^{-1}$, after the subtraction of the systemic velocity of the galaxy. Bottom panels: velocity dispersion maps for the H$\alpha$ (left) and [N\,{\sc ii}] (right) emission lines, corrected for the instrumental broadening. The color bars show the $\sigma$ values in units of km\,s$^{-1}$. The central cross in all panels marks the position of the nucleus and grey regions in the flux and $\sigma$ maps and white regions in the $V_{LOS}$ maps represent masked locations, where the signal-to-noise was not high enough to obtain reliable fits to the emission line profiles or locations with no line detection. The green contours in the velocity dispersion map of H$\alpha$ line are  from the 3~cm radio image.  The grey asterisks labelled as OH1 and OH2 mark the locations where the maser emission has been detected.} 

    \label{fig:all}
    \end{flushleft}
\end{figure}

The top panels of Figure \ref{fig:all} present the flux distributions in the narrow components of H$\alpha$ (left) and [N\,{\sc ii}]$\lambda$6583 (right) emission-lines. The color bars show the flux in logarithmic units of erg s$^{-1}$ cm$^{2}$ and the grey regions represent masked locations where the uncertainty in the flux is larger than 30\%. The central cross marks the location of the nucleus, defined as the position of the peak of the flux distribution of the broad H$\alpha$ component and is labelled with the letter N in Fig.~\ref{fig:all}. The two lines show similar flux distributions, with the emission within the inner $\sim1^{\prime\prime}$ being elongated in the northeast--southwest direction. At least three extranuclear knots of emission, labelled A, B and C in Fig.~\ref{fig:all}, are observed in the H$\alpha$ flux map, at (x,y) angular distances relative to the nucleus of ($-$0\farcs6,$-$2\farcs2), ($-$1\farcs0,1\farcs8) and (1\farcs0,$-$0\farcs6). We note that these knots are not observed in the HST [N\,{\sc ii}]+H$\alpha$ image.

\subsection{Gas velocity fields and velocity dispersion maps}

The central panels of Fig.\,\ref{fig:all} present the line-of-sight velocity ($V_{LOS}$) fields for the H$\alpha$ (left) and [N\,{\sc ii}]$\lambda$6583 (right) emission-lines. White regions were masked following the same criteria used for the flux maps, described above. The systemic velocity of $V_{s}$=37\,947 km\,s$^{-1}$ was subtracted from the observed velocities, as derived from the modelling of the H$\alpha$ velocity field with a rotating disk model (see Sec.~\ref{gas_kin}).

The velocity fields derived from H$\alpha$ and [N\,{\sc ii}] emission lines are similar, presenting  redshifts to the west and blueshifts to the east of the nucleus with velocities reaching up to 200~km\,s$^{-1}$.  The zero velocity line presents an $S$ shape and values close to the systemic velocity are observed at $\sim 2^{\prime\prime}$ to the north of the nucleus.

%\subsection{Velocity dispersion maps}
The bottom panels of Fig.\,\ref{fig:all} show the velocity dispersion ($\sigma$) maps for H$\alpha$ (left) and [N\,{\sc ii}]$\lambda$6583 (right), corrected for the instrumental broadening. These maps show values ranging from $\sim40$ to 150~\kms, with the highest values observed co-spatially with the $S$ shaped zero velocity curve of the $V_{LOS}$ maps (central panels). In addition, some high $\sigma$ values are seen for H$\alpha$ at 1\farcs5 northeast of the nucleus, where the velocity fields present the highest blueshifts. The smallest values of $\sigma$ are co-spatial with the extranuclear knots of emission observed in the emission line flux distributions in the top panels of Fig.\,\ref{fig:all}.

\section{Discussion}

In this work, we present for the first time high quality HST images of IRASF23199+0123. These images allowed us to identify that this system is indeed composed by two members and discern tidal structures.  Thus, this is the first time that this system has been established to be an interacting galaxy pair. 

\subsection{Gas excitation and diagnostic diagrams}

The [N\,{\sc ii}]$\lambda$6583/H$\alpha$ flux ratio can be used to map the gas excitation \citep[e.g.][]{baldwin81,cid10}, with values [N\,{\sc ii}]$\lambda$6583/H$\alpha\leq$1 corresponding to gas ionized by young stars, while larger values correspond to gas excited by a central AGN or shocks \citep[e.g.][]{sb07,cid11}. 

\begin{figure}
	\includegraphics[width=\columnwidth]{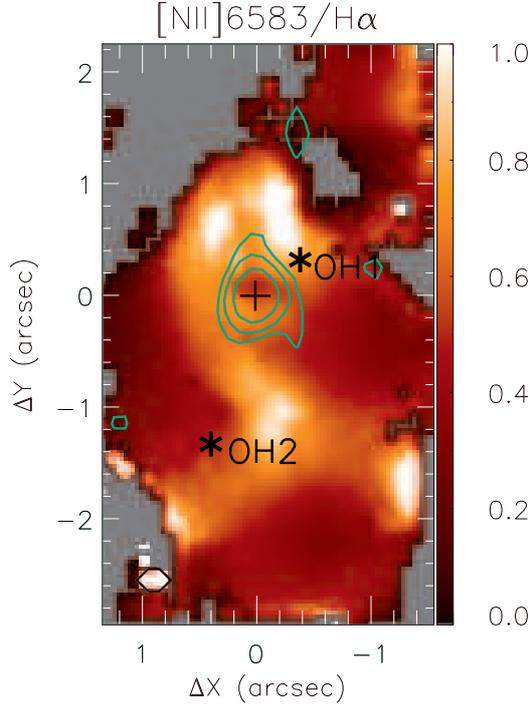}
    \caption{[N\,{\sc ii}]/H$\alpha$ flux ratio map for IRAS23199E. Grey regions correspond to masked locations where no reliable measurements are available. The green contours in the map are from the 3~cm radio image. The black asterisks labelled as OH1 and OH2 mark the position of the maser emission}.
    \label{fig:razaoNIIporHalpha}
\end{figure}

In Figure~\ref{fig:razaoNIIporHalpha} we present the  [N\,{\sc ii}]$\lambda$6583/H$\alpha$ ratio map for IRAS23199E. Grey regions correspond to locations masked due to poor fits. The map shows values ranging from $\sim$~0.2 to 1, with the highest values observed approximately coincident with the region with the highest $\sigma$ values (bottom panels of Fig.~\ref{fig:all}), while the lowest values are observed in the knots of enhanced H$\alpha$ emission (top-left panel of Fig.~\ref{fig:all}). We interpret these latter locations as complexes of star forming regions. 

The fact that the locations with the highest [N\,{\sc ii}]$\lambda$6583/H$\alpha$ ratios are co-spatial with the highest $\sigma$ values suggests that shocks contribute to the gas excitation. We have extracted a spectrum within a circular aperture of 0\farcs4 radius in one of these locations, labelled E in Fig.~\ref{fig:all}. We have measured the line ratios and equivalent width of H$\alpha$ from this spectrum, to plot this region in the diagnostic diagram WHAN \citep{cid10}. This diagram has been proposed as an alternative to the BPT diagrams of \citet{baldwin81}, and is a plot of the the H$\alpha$ equivalent width against the  [N\,{\sc ii}]$\lambda$6583/H$\alpha$ flux ratio. While BPT diagrams need four emission lines to separate the regions ionized by AGN or starburst, the WHAN diagram requires just H$\alpha$ and \nii. The WHAN diagram enables a separation between Starbusts, Seyfert galaxies (sAGN) and low-luminosity AGNs (wAGN). It is also possible to separate wAGN population, where the emitting gas is excited by a central AGN from Retired Galaxies (RG), where the gas emission may be due to excitation by hot, evolved (post-asymptotic giant branch - post-AGB) stars, in which case the H$\alpha$ equivalent width is smaller than 3\AA \citep[e.g.][]{belfiore16,brum17}. 

\begin{figure*}
	\includegraphics[width=0.9\columnwidth]{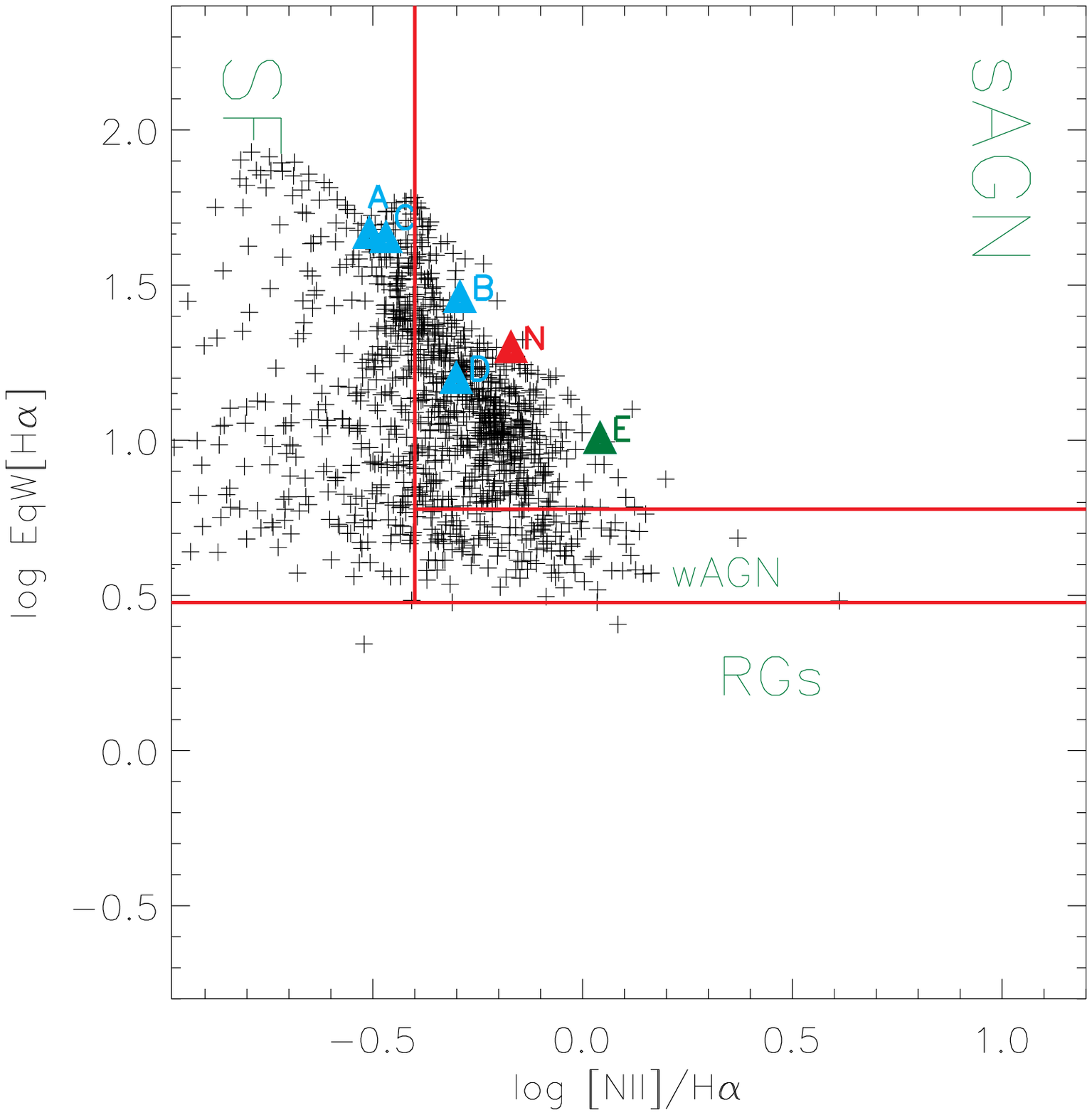}
	\includegraphics[width=0.9\columnwidth]{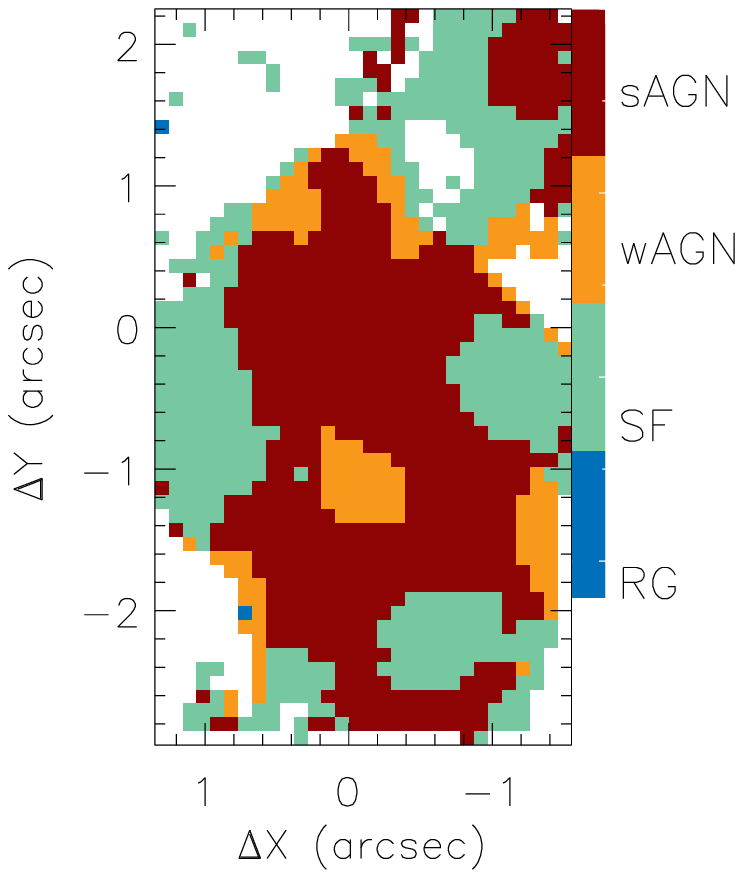}
    \caption{Left: the WHAN diagram for IRAS23199E showing the different excitation regions \citep{cid10}. 
%The vertical red line at log[N\,{\sc ii}]/H$_{\alpha}=-0.40$ refers to the division between star forming and AGN classes. The division at logEqW[H$\alpha$]=0.8 represents the separation between Seyferts and LINERs. 
Each cross corresponds to an individual spaxel of the IFU datacube; the red triangle represents the nucleus; blue  triangles are for regions A, B, C and D identified in the H$\alpha$ flux map of Fig.~\ref{fig:all} and the green triangle corresponds to the region E identified in the [N\,{\sc ii}] $\sigma$ map of Fig.~\ref{fig:all}. Right: Excitation map identifying the regions within the FoV corresponding to different excitation mechanisms: strong AGNS (sAGN), weak AGN (wAGN), Starforming (SF) and Retired Galaxy (RG). }
    \label{fig:whan}
\end{figure*}

The WHAN diagram for IRAS23199E is shown in the left panel of Figure \ref{fig:whan}, while the right panel of this figure shows the corresponding excitation map, with the distinct excitation classes identified. The red triangle represents the nucleus, as obtained from the fitting of the line profiles from an integrated nuclear spectrum within 0\farcs4 radius. Although the H$\alpha$ broad component is seen to up to 0\farcs8 from the nucleus -- due to the wings of the PSF, the smaller aperture of 0\farcs4 was chosen here to focus on the AGN emission, most of which is contained within this aperture (the PSF FWHM  measured from field stars flux distribution is 0\farcs85). The blue triangles correspond to the extra-nuclear locations identified in Fig.~\ref{fig:all}. The green triangle corresponds to the spectrum from region E described above where there is an enhancement of both the line ratio and the velocity dispersion $\sigma$. 
%, used to construct the BPT diagram of Fig.~\ref{fig:all}.
All points are located within the region expected for emission of gas excited by an AGN or Starburst. In particular, the point corresponding to the nucleus is clearly located within the sAGN region (strong AGN), indicating that the nuclear emission originates in a Seyfert type AGN, in agreement with the presence of the broad H$\alpha$ component. In addition to the nucleus, region E can also be classified as sAGN. At this location, the [N\,{\sc ii}]$\lambda$6583/H$\alpha$ ratio is even larger than for the nucleus, possibly that shocks contribute to the gas excitation.. The blue triangles corresponding to the extra-nuclear regions identified in Fig.\,\ref{fig:all} are located very close to the line that separates starburst from sAGN excitation, indicating that both the central AGN and star forming regions contribute to the gas excitation at these locations.
%, in good agreement with the BPT diagram (Fig.~\ref{fig:bpt}).

%In order to further investigate this possibility, we constructed the plot of [N\,{\sc ii}]$\lambda$6583/H$\alpha$ vs. $\sigma$ shown in Figure~\ref{fig:niihasigma}. 

%that  [N\,{\sc ii}]$\lambda$6583/H$\alpha$ is correlated $\sigma$. 
%We used the  $\sc idl$ routine $\sc r_-correlate.pro$ to compute the Spearman's correlation coefficient ($\rho$) between [N\,{\sc ii}]/H$\alpha$ and $\sigma$, which results $\rho=0.68$. 

 We note that there is a general trend for higher [N\,{\sc ii}]$\lambda$6583/H$\alpha$ to be associated with larger $\sigma$ values. Some of these regions surround the contours of the 3\,cm radio emission, as shown in Fig.\,\ref{fig:razaoNIIporHalpha}. We thus attribute the gas emission at high-$\sigma$ and [N\,{\sc ii}]/H$\alpha$ ratio as at least being partly produced by excitation of the gas by shocks associated with produced by the radio-emitting plasma. There are also some regions to the south of the nucleus where there is no radio emission but there is still enhancement of the $\sigma$ and line ratios, which we attribute to the presence of additional perturbations possibly due to the interaction between the two galaxies in IRASF23199+0123.

%We have plotted the ratio [N\,{\sc ii}]/\Ha against the velocity dispersion obtained from the width of the [N\,{\sc ii}] $\lambda$6584 emission line, $\sigma_{[N\,{\sc ii}]}$ as one can see in figure \ref{fig:niihasigma}. It was an attempt in order to investigate the excitation mechanism that ionizes the gas. There is a correlation between [N\,{\sc ii}]/\Ha~ and $\sigma_{[N\,{\sc ii}]}$. It can indicate that gas excitation is correlated to its kinematics. In addiction it suggests that part of the ionization is produced by shocks (artigo da Thaisa). The presence of shocks could also be notice in figure \ref{fig:all}, which evince the highest values of velocity dispersion at the nucleus.

%%%%%%%%%%%%%%%%%%%%%%%%%%%%%%%%%%%%%%%%%%%%%%%%%%%%%%%%%%%%%%%%%%%%%%%%%%%%%%%%%%%%%%%%%%%%%%%%%%%%%%%%%%%%%%%%%%%%%

\subsection{Star forming regions}
As discussed in previous sections, we identify several knots of enhanced H$\alpha$ emission, associated to star forming regions. These knots are labelled A, B and C in Fig. \ref{fig:all}. In order to characterize the star formation at these locations, we use the integrated fluxes within the circular apertures shown in Fig.~\ref{fig:all}. These fluxes have been used to estimate physical properties of the star forming regions that are listed in Table~\ref{tabela}.
In order to estimate the mass of ionized gas we used \citep{peterson97}:

\begin{equation}
    \frac{M}{\rm M_{\odot}} \approx 2.3 \times 10^{5} \frac{L_{41}(H_{\alpha})}{n_{3}^{2}}, 
\end{equation}
where L$_{41}$(H$\alpha$) is the H$\alpha$ luminosity in units of 10$^{41}$ erg s$^{-1}$ and n$_{3}$ is the electron density ($N_e$) in units of 10$^3$\,cm$^{-3}$. We have assumed $N_e=300$ \,cm$^{-3}$, which is the mean value of electron density of circumnuclear star formation regions derived from the [S\,{\sc ii}]$\lambda$\,6717/$\lambda$\,6731 intensity ratio \citep{diaz07,oli08}.  The values of the mass of ionized gas for each star forming complex are in the range (1.78 -- 4.08) $\times$ 10$^{5}$ M$_{\odot}$.

%\textcolor{magenta}{Atualizar os numeros que dependem da densidade eletronica. Note que substitui n3=0.1 por n3=0.3. refaz as contas para este novo valor. Atualizar tb a tabela.}

We estimated the rate of ionizing photons $Q$[H$^{+}$] and star formation rate (SFR) under the assumption of a continuous star formation regime.
The rate of ionizing photons for each star formation region was derived using \citet{osterbrock}:

\begin{equation}
 Q[H^{+}]=\frac{\alpha_{B}L_{H_{\alpha}}}{\alpha^{EFF}_{H_{\alpha}}h\nu_{H_{\alpha}}},
\end{equation}
where $\alpha_{B}$ is the hydrogen recombination coefficient to all energy levels above the ground level, $\alpha^{EFF}_{H_{\alpha}}$ is the effective recombination coefficient for H$\alpha$, $h$ is the Planck's constant and $\nu_{H_{\alpha}}$ is the frequency of the H$\alpha$ line. Using $\alpha_{B}$=2.59$\times$ 10$^{13}$ cm$^{3}$s$^{-1}$ and $\alpha^{EFF}_{H_{\alpha}}$=1.17 $\times$ 10$^{-17}$cm$^{3}$s$^{-1}$ \citep{osterbrock} we obtain:

\begin{equation}
    \left(\frac{Q[H^{+}]}{s^{-1}}\right)=1.03\times10^{12}\left(\frac{L_{H_{\alpha}}}{s^{-1}}\right).
\end{equation}

The star formation rate (SFR) was computed using the following relation \citep{kennicutt98}:
\begin{equation}
\frac{SFR}{\rm M_\odot\,yr^{-1}}=7.9\times10^{-42}\,\frac{L_{\rm H_\alpha}}{\rm  erg\, s^{-1}} 
\end{equation}

SFRs  derived for the star formation regions of IRAS23199E are in the range 0.05 -- 0.12 M$_{\odot}$yr$^{-1}$, consistent with a moderate star-forming regime. These SFRs fall within the range observed for circumnuclear star forming regions in nearby Seyfert galaxies, derived using optical \citep{oli08} and near-infrared \citep{Falcon2014,riffel2016,hennig2017} emission-lines and are consistent with the average value of $SFR=0.14$ M$_{\odot}$ yr$^{-1}$ for a sample of 385 galaxies.

The values of the ionizing photons rate  are in the range $\log$ Q[H$^{+}$]$=$(51.87 - 52.23) s$^{-1}$ and are in agreement with previous reported values for circumnuclear star forming regions in nearby galaxies \citep[e.g.][]{Wold06,Galliano08,riffel09,riffel2016}.

The masses of ionized gas derived for the star-forming complexes in IRASF23199E are in the range of (1.78 - 4.08)$\times10^{5}$ M$_{\odot}$, and agree with those previously obtained for star forming regions surrounding Seyfert nuclei \citep[e.g.][]{riffel2016}. 

 We can also use the Far-Infrared luminosity of galaxy to calculate the SFR \citep{kennicutt98}:

\begin{equation}
SFR{\rm (M_{\odot} yr^{-1})}=4.5 \times10^{-44} L_{FIR}{\rm( erg s^{-1})}
\end{equation}
where $L_{FIR}$ is the IR luminosity integrated over the mid- and far-IR spectrum (8 - 1000 $\mu$m). Using $L_{\rm FIR}=1.35\times10^{11}$ erg\,s$^{-1}$ \citep{Darling2006}, we obtain SFR$\approx$61 ${\rm M_{\odot}}$yr$^{-1}$. This value is much higher than those obtained for each star-forming region (shown in Tab.~\ref{tabela}) and suggests that most of the FIR luminosity may not be due to star formation, but to the  AGN, or  most of the star formation is embedded in dust.

%These larger values of ionized gas masses obtained for IRASF23199E may be due to the larger aperture used in this work.

\begin{table*}
	\centering
	\caption{Physical properties of the star formation regions in IRAS23199E.}
	\label{tabela}
	\begin{tabular}{cccccc} % four columns, alignment for each
		\hline
		Region & L$_{H_{\alpha}}$ (10$^{41}$ erg s$^{-1}$) & EqW[H$\alpha$] & M (10$^{5}$M$_{\odot}$) & log Q[H$^{+}$] (s$^{-1}$) & SFR (M$_{\odot}$ yr$^{-1}$)\\
		\hline
		A & 0.12 & 44.41 & 3.06 & 52.10 & 0.09\\
		B & 0.07 & 28.41 & 1.78 & 51.87 & 0.05\\
		C & 0.16 & 45.84 & 4.08 & 52.23 & 0.12\\
		\hline
	\end{tabular}
\end{table*}

%%%%%%%%%%%%%%%%%%%%%%%%%%%%%%%%%%%%%%%%%%%%%%%%%%%%%%%%%%%%%%%%%%%%%%%%%%%%%%%%%%%%%%%%%%%%%%%%%%%%%%%%%%%%%%%%%%%%%

%%%%%%%%%%%%%%%%%%%%%%%%%%%%%%%%%%%%%%%%%%%%%%%%%%%%%%%%%%%%%%%%%%%%%%%%%%%%%%%%%%%%%%%%%%%%%%%%%%%%%%%%%%%%%%%%%%%%%
\subsection{Gas kinematics} \label{gas_kin}

\begin{figure*}
	\includegraphics[width=1.8\columnwidth]{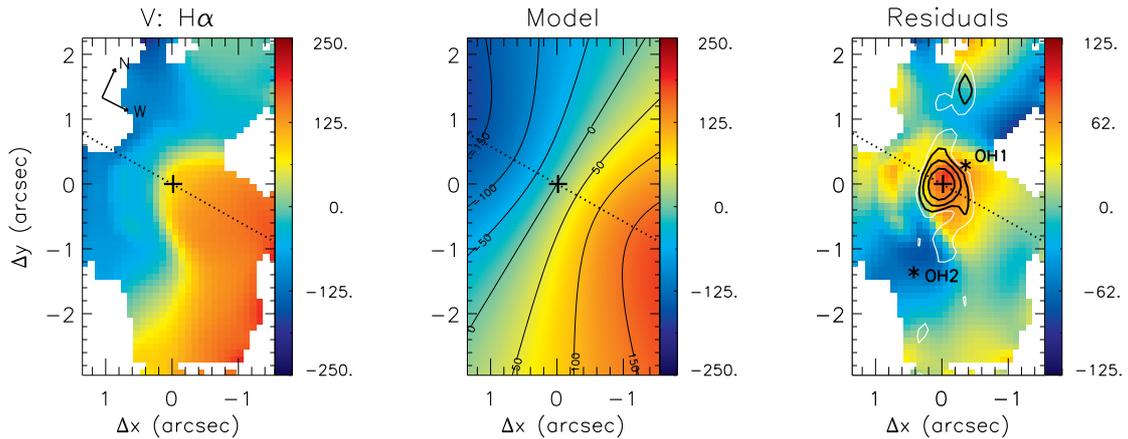}
	\caption{Observed H$\alpha$ velocity field (left), rotating disk model (center) and residual between the two (right). 
	%The near and the far side of the galaxy disk are indicated in the left panel,
	The central cross marks the position of the nucleus, the white regions are masked locations where we were not able to fit the emission line profiles and the dotted lines represent the orientation of the line of nodes. The black contours in the residual map are  from the 3~cm radio image with the same flux levels as shown in Fig.~\ref{fig:radiocontours} and the white contours show  radio contours at the 1.5$\sigma$ level.  The black asterisks labelled OH1 and OH2 mark the position of the maser sources.
}
	\label{fig:rotationmodel}
\end{figure*}

The velocity fields shown in Fig.~\ref{fig:all} are complex, but they suggest the presence of a rotation pattern with the line of nodes oriented approximately along the east-west direction. 
In order to describe analytically this behaviour, we used a simple rotation model \citep{kruit78,bertola91},which  assumes that the gas moves in circular orbits in the plane of the galaxy, within a central gravitational potential. In this model, the rotation velocity field is given by:

\[ 
 V_{mod}(R,\psi)=v_{s}+ 
\]
\begin{equation}
    \frac{AR\cos(\psi-\psi_{0})\sin(i){\cos^{p}(i)}}{\{R^{2}[\sin^{2}(\psi-\psi_{0})+\cos^{2}(i)\cos^{2}(\psi-\psi_{0})]+{c_{0}}^{2}\cos^{2}(i)\}^{\frac{p}{2}}}
    \end{equation}
where $R$ and $\psi$ are the coordinates of each pixel in the plane of the sky, $v_{s}$ is the systemic velocity of the galaxy, $A$ is the velocity amplitude, $\psi_{0}$ is the major axis position angle, $i$ is the disc inclination relative to the plane of the sky ($i=0$ for face-on disc), $p$ is a model fitting parameter (for {\it p}\,=\,1 the rotation curve at large radii is asymptotically flat
while for {\it p}\,=\,3/2 the system has a finite mass) and $c_{0}$ is a concentration parameter,  defined as the radius where the rotation curve reaches 70\,\% of its velocity amplitude.

%Perform this model provide us information about physical parameters from the system, like the systemic velocity and the 
%orientation of the line of nodes. Moreover, the residual map (difference between the analytic model and the observed field) allow us to detect some deviations from the usual rotation pattern as well as identify possible outflows and/or outflows.

The [N\,{\sc ii}] and H$\alpha$ emission-lines present similar velocity fields (Fig.~\ref{fig:all}), so we have chosen the H$\alpha$ velocity field to perform the fit, as this line is stronger than [N\,{\sc ii}]$\lambda6583$ at most locations.

The observed velocities were fitted with the equation above using the MPFITFUN routine \citep{mark09} in IDL\footnote{\rm $http://www.harrisgeospatial.com/ProductsandSolutions/Geospatial$\\$Products/IDL.aspx$}, that performs a non-linear least-squares fit, after initial guesses for the parameters. During the fit, the position of the kinematical center was kept fixed to the position of the peak of the flux distribution of the broad H$\alpha$ component, adopted as the location of the nucleus of the galaxy and the parameter $p$ was kept fixed at $p=1.5$, as done in previous works \citep[e.g.][]{brum17}.

The resulting best fit model is shown in the central panel of Figure~\ref{fig:rotationmodel} and its parameters are $A$=349$\pm$26 km s$^{-1}$, $v_{s}$=37\,947$\pm$2 km s$^{-1}$ (corrected for the heliocentric rest frame), $\psi_{0}$=95$\pm$2$^{\circ}$, $c_{0}$=1\farcs6$\pm$0\farcs1, i=41$^\circ\pm$6$^{\circ}$. The systemic velocity of the galaxy can be compared with previous measurements.   \citet{lawrence99} described the construction of the QDOT survey, which consists of infrared properties and redshifts of an all-sky sample of 2387 IRAS galaxies. They obtained a systemic velocity for IRASF23199+0123 of 40\,981 \kms.
On the other hand, \citet{Darling2006} performed an optical spectrophotometric study of resolved spectra of multiple nuclei merging systems that hosts OHM sources and obtained  $v_{s}$=40\,679$\pm$2 km s$^{-1}$.  We note that the systemic velocity derived here is smaller than those obtained in previous studies.  We speculate that this difference may be due to the fact that previous works possible have observed the western nucleus, which is brighter,  since at that time it was not known that IRASF23199+0123 is composed by two members. 

Besides the disk rotation model, Fig.~\ref{fig:rotationmodel} shows also the observed H$\alpha$ velocity field in the left panel and the residual map between the observed velocities and the model in the right panel. The residual velocity map shows values much smaller than the observed velocity amplitude, but residuals of up to 100~\kms\ are present at some locations. Blueshifts are seen to the south and north of the nucleus, while redshifts are observed at the nucleus and its surroundings. 

In order to investigate the origin of the velocity residuals, we have overlaid the contours from the 3\,cm radio image on the residual map shown in the right panel of Fig.~\ref{fig:rotationmodel}. In this figure, we have plotted also the radio contours at the 1.5$\sigma$ level (in white), as these levels show that faint radio emission is elongated towards the region where  blueshifts are observed to the north of the nucleus. At the 3$\sigma$ level there is just a hint of this elongation.
% while the redshifts are seen at the same orientation where the radio contours show an elongated structure. These close association of the radio and velocity residual maps suggests that the 
This apparent association of the extended radio emission (although at faint levels) with the blueshifts in the map of velocity residuals suggests that the radio-emitting plasma may play a role in the gas kinematics and a possible interpretation is that the blueshifted residuals are due gas pushed away from the nucleus  by a radio jet. Other indications of the interaction between the radio plasma and the emitting gas are the higher velocity dispersion and \nii/H$\alpha$ values surrounding the radio structures, as seen in Figs.~\ref{fig:all} and \ref{fig:razaoNIIporHalpha}. Similar results have been found for other galaxies and interpreted as originating from the interaction of the radio jet with the ambient gas \citep[e.g.][]{riffel06,riffel15}.

We speculate that the observed blueshifts originate in outflows along a bi-cone oriented in the north-south direction with its axis approximately in the plane of the sky. The blueshifts would come from the front walls (nearside) of the cones to both sides of the nucleus, while the redshift from the back (farside) walls  are not observed probably due to obscuration. 
The redshifts  that are observed surrounding  the nucleus could be due to inflows towards the nucleus, probably due to gas motions associated with the interaction between the two galaxies of the pair.

Finally, it is interesting to note that the OH masers are observed in the vicinity of the active nucleus of IRASF23199E, close to regions of enhanced velocity dispersion and [N\,{\sc ii}]/H$\alpha$ ratio in the emitting gas. This suggests that the maser sources are associated with the AGN, perhaps produced in gas compressed in an interaction with expanding radio plasma. We also notice that the redshifted maser source (OH1), is located in a region with redshifted residuals, while the blueshifted maser source (OH2),  is located in a region of blueshifted residuals, suggesting that  the OH2  source is participating in the outflow.

\subsection{The nature of the nuclear emission} \label{broad}

\citet{Darling2006} presented a study of the optical properties of the Arecibo Observatory OH megamaser survey sample, with the aim of investigating the types of nuclear nuclear environments that produce OH megamasers. They determined that IRASF23199+0123 harbours a Seyfert 2 nucleus, based on an optical spectrum but the spectra used for the classification includes both nuclei, thus it was not possible to reveal the nature of each nucleus separately.

With our data we have discovered that IRAS23199E harbours a Seyfert 1 nucleus, as we clearly detected a broad unresolved H$\alpha$ component at the nucleus and the WHAN diagram is consistent with Seyfert-like gas excitation there. The double-peaked nature of this broad  component suggests it is due to unresolved disk rotation in the BLR, as many recent studies have supported a flattened geometry for this region \citep{sb17}. In the NLR, jet-cloud interaction may give rise to double-peaked lines \citep[e.g.][]{capetti99,odea02,preeti17} and another possibility is that some interaction with the radio emission may be happening in the BLR is this particular case. The discrepancy between Darling's classification and ours can be understood if \citet{Darling2006} observed the western nucleus, which appears brighter in our HST image and in the Sloan Digital Sky Survey (SDSS) image \citep{bundy15,sdss}.

% The observation of the broad H$\alpha$ component can be reconciled with the speculative scenario proposed in the previous section of an inflow in the vicinity of the nucleus, if we are seing the broad H$\alpha$ component through holes in the dusty torus postulated by the Unified Model of AGNs. This is compatible with the recent studies that show that the torus is most probably composed by large number of small clouds as modeled by \citep[e.g.][]{nenkova08}, and raises the possibility also that the broad BLR was hidden by clouds during the observations by \citet{Darling2006} and then uncovered by the motion of the clouds in our more recent observation.

%Similar results have been previously observed \citep[e.g.][]{riffel14,Lena15} and are supported by recent modelling  of accretion disk winds and outflowing torus  \citep{li13,honig13,elitzur12,ivezic10,mor09,nenkova08,elitzur06}. 
%As a speculation we can estimate the angular displacement of a cloud of the dusty torus during the time between the observations used by \citet{Darling2006} and ours. They used data obtained in 2000 and thus the time between theirs and ours observations is about 13 yr, as our data were obtained in 2013. Assuming a cloud located at the inner region of the dusty torus (at 0.1\,pc) subjected only to the gravitational potential of a central SMBH of mass of 10$^7$\,M$_\odot$, and considering a rigid body rotation, we obtain that the cloud would move about 6$^\circ$ during the period between Darling's and our observations.  

We estimate the mass of the central black hole using the empirical relation given by \citet{greene05}:

\[ 
  \frac{M_{BH}}{\rm M_ \odot}=(2.0^{+0.4}_{-0.3}) \times 10^{6}
\]
\begin{equation}
    \left(\frac{L_{H_{\alpha}}}{10^{42}{\rm erg s^{-1}}}\right)^{0.55\pm 0.02}\left(\frac{FWHM_{H_{\alpha}}}{10^{3}{\rm kms^{-1}}}\right)^{2.06\pm 0.06}
\end{equation}
where $M_{\rm BH}$ is the black hole mass, $L_{\rm H_{\alpha}}$ is the luminosity and $FWHM_{\rm H_{\alpha}}$ is the full width at half maximum of the broad component. The luminosity was calculated as the sum of the luminosities of both components, resulting in  $L_{\rm H_{\alpha}}\approx 1.8 \times10^{41}$ erg\,s$^{-1}$. 
 We obtained the $FWHM_{\rm H_{\alpha}}$ of 2170 \kms directly from the observed profile. 
%\begin{equation}
%    FWHM_{\rm H_{\alpha}}=\frac{FWHM_{1}F_{1}+FWHM_{2}F_{2}}{F_{1}+F_{2}} \times 2
%\end{equation}
%where $FWHM_{1}$ and $FWHM_{2}$ are the FWHMs of the blue and red components, respectively. $F_{1}$ and $F_{2}$ are the fluxes of each component. 
 Using these values we have estimated a black hole mass of $3.8^{+0.3}_{-0.2}\times10^{6}$\,M$_{\odot}$.
%%%%%%%%%%%%%%%%%%%%%%%%%%%%%%%%%%%%%%%%%%%%%%%%%%%%%%%%%%%%%%%%%%%%%%%%%%%%%%%%%%%%%%%%%%%%%%%%%%%%%%%%%%%%%%%%%%%%

\section{Conclusions}

We present a multi-wavelength study of the OH megamaser galaxy IRASF23199+0123 using the HST, VLA and Gemini North telescope.  Our HST images show that this system is an interacting pair of galaxies and used integral field spectroscopic data obtained with the Gemini GMOS-IFU to observe its eastern nucleus, which we call IRASF23199E. Our observations cover the inner 9.5$\times$13~kpc$^2$ of the galaxy at a spatial resolution of 2.3~kpc and velocity resolution of $\sim70$\kms.  Our main conclusions are:

\begin{itemize}

\item We show that IRASF23199+0123 is an interacting pair with a tidal tail connecting the two galaxies and detect two OH maser sources associated to the eastern member.

\item Both nuclei present extended radio emission at 3 and 20~cm, with intensity peaks at the each nucleus. The 20~cm radio emission of the eastern nucleus is elongated in the direction of the most extended emission in the HST continuum image (northeast-southwest), while in the western nucleus the 20~cm radio emission is tilted by about 45$^{\circ}$ eastwards relative to the orientation of the most extended continuum emission.  In the better spatially resolved 3~cm observations, some elongation is observed at low brightness level towards the north in the eastern nucleus.

\item One of the main results of this paper is the discovery of a Seyfert~1 nucleus in IRASF23199E, via the detection of an unresolved broad (FWHM$\approx$2\,200\,km\,s$^{-1}$) double peaked component in the H$\alpha$ emission-line from the BLR. This is important in regard to the OH maser emission, because the two masing sources are detected in this galaxy which hosts an AGN, rather than the other member of the pair. In addition, the masing sources are observed in the vicinity of enhanced velocity dispersion and higher line ratios, suggesting that they are associated with shocks driven by AGN outflows. The blue and redshifted maser sources are associated with the blue and redshifted ionized gas velocity residuals. This combination of evidence from HST images, VLA line spectroscopy and IFU spectroscopy strongly indicates that in this system, the OH megamaser sources are associated with the AGN, rather than star formation.

\item Using the width and luminosity of the broad H$\alpha$ profile, we estimate a mass of M$_{BH}$= 3.8$^{+0.3}_{-0.2}$ $\times$ 10$^{6}$M$_{\odot}$ for the central SMBH.

\item The comparison between the HST [N\,{\sc ii}]+H$\alpha$ image and GMOS-IFU emission-line flux distributions of IRASF23199E shows that they are similar, being more elongated in the northeast-southwest direction, following the continuum emission. In addition, the GMOS H$\alpha$ flux map reveals the presence of three extra-nuclear knots, attributed to star forming regions.

\item From the measurement of the H$\alpha$ fluxes from the star forming regions, we obtain: (1) star formation rates in the range (0.05 -- 0.12) M$_{\odot}$ yr$^{-1}$; (2) ionized gas content in the range (1.8 -- 4.1) $\times$ 10$^{5}$ M$_{\odot}$; and (3) ionized photons rate log Q[H$^{+}$]=51.9 to 52.2 s$^{-1}$. From the FIR luminosity we obtain obtain SFR$\approx$61 ${\rm M_{\odot}}$yr$^{-1}$ suggesting that most of the FIR luminosity may be due to the  AGN or  due to star formation embedded in dust.

\item The [N\,{\sc ii}]$\lambda$6583/H$\alpha$ flux ratio map of IRAS23199E presents the highest values coincident with the regions with highest $\sigma$ values, which surround the radio source and suggests that the radio-emitting plasma interacts with the ambient gas via shocks that seem to play a role in the gas excitation. The lowest [N\,{\sc ii}]$\lambda$6583/H$\alpha$ values are co-spatial with the star forming regions detected in the H$\alpha$ emission.
 
 \item The velocity fields of IRASF23199E show a disturbed rotation pattern with the line of nodes oriented along $PA=95^\circ$, as derived by the fit of the H$\alpha$ velocities with a rotation disk model. The residuals between the observed and modelled velocity field combined with the velocity dispersion maps suggest the presence of non circular motions, possibly due to outflows from the nucleus along the directions north and south and inflows towards the nucleus in its vicinity.

%\item The outflows are seen as blueshifts in the residual map, seen to the northwest and to the southeast, coincident with the orientation of the radio structures. These residuals are interpreted as being originated within a bi-cone with central axis approximately perpendicular to us, so that the blueshifts are originated from the front shells of the cones. 

%\item The residual map shows also redshifts at the nucleus to the west of it, approximately along the major axis of the galaxy, co-spatial with a radio extended structure. We speculate two scenarios in such this kinematic component would be consistent with gas outflows from the central AGN or inflows towards it. 

%\item {\bf Indications} of the interaction of the {\bf radio-emitting plasma} with the ambient gas are seen as enhancements in the velocity dispersion and [N\,{\sc ii}]/H$\alpha$ values at some locations, surrounding the radio structures.

\end{itemize}

\section*{Acknowledgements}
We thank an anonymous referee for useful suggestions which helped to improve the paper. 
This work is based on observations obtained at the Gemini Observatory, 
which is operated by the Association of Universities for Research in Astronomy, Inc., under a cooperative agreement with the 
NSF on behalf of the Gemini partnership: the National Science Foundation (United States), the Science and Technology 
Facilities Council (United Kingdom), the National Research Council (Canada), CONICYT (Chile), the Australian Research 
Council (Australia), Minist\'erio da Ci\^encia e Tecnologia (Brazil) and south-eastCYT (Argentina). 
This research has made use of the NASA/IPAC Extragalactic Database (NED) which is operated by the Jet Propulsion Laboratory, California Institute of Technology, under contract with the National Aeronautics and Space Administration.
The National Radio Astronomy Observatory is a facility of the National Science Foundation operated under cooperative agreement by Associated Universities, Inc.
We acknowledge the usage of the HyperLeda database (http://leda.univ-lyon1.fr).
C. H. thanks for CAPES financial support.  R.A.R. acknowledges support from FAPERGS and CNPq.

%%%%%%%%%%%%%%%%%%%%%%%%%%%%%%%%%%%%%%%%%%%%%%%%%%



%\bibliographystyle{mnras}


%%%%%%%%%%%%%%%%%%%%%%%%%%%%%%%%%%%%%%%%%%%%%%%%%%

%%%%%%%%%%%%%%%%% APPENDICES %%%%%%%%%%%%%%%%%%%%%

%\appendix



%%%%%%%%%%%%%%%%%%%%%%%%%%%%%%%%%%%%%%%%%%%%%%%%%% 0.88 resolução angular.

% Don't change these lines
\bsp	% typesetting comment
\label{lastpage}
\end{document}